\documentclass[10pt,preprint]{aastex}
\usepackage{apjfonts}

\newcommand{\pixrate}{\langle\Gamma_{\rm p}\rangle}
\newcommand{\pixraten}{\pixrate_{\rm n}}
\newcommand{\pixratef}{\pixrate_{\rm f}}
\newcommand{\te}{t_{\rm E}}
\newcommand{\timeratio}{\langle\te\rangle_{\rm f}/\langle\te\rangle_{\rm n}}
\newcommand{\sm}{\mbox{M}_{\odot}}

\newcommand{\Nf}{N_{\rm f}}
\newcommand{\Nn}{N_{\rm n}}
\newcommand{\Nt}{N_{\rm t}}
\newcommand{\Nc}{N_{\rm c}}
\newcommand{\fc}{f_{\rm c}}
\newcommand{\half}{\frac{1}{2}}

\begin{document}
\title
{Theory of pixel lensing towards M31 --\\ 
II. The velocity anisotropy and flattening of the MACHO distribution}
\author{E.~Kerins\altaffilmark{1}, J.~An\altaffilmark{2},
N.~W.~Evans\altaffilmark{2}, P.~Baillon\altaffilmark{3},
B.J.~Carr\altaffilmark{4}, Y.~Giraud-H{\'e}raud\altaffilmark{5},
A.~Gould\altaffilmark{6}, P.~Hewett\altaffilmark{2},
J. Kaplan\altaffilmark{5}, S.~Paulin-Henriksson\altaffilmark{5},
S.J.~Smartt\altaffilmark{2}, Y.~Tsapras\altaffilmark{4},
D.~Valls-Gabaud\altaffilmark{7}\\
(The POINT--AGAPE Collaboration)}

\altaffiltext{1}{Astrophysics Research Institute, Liverpool John
Moores University, Egerton Wharf, Birkenhead, CH41 1LD, UK}
\altaffiltext{2}{Institute of Astronomy, University of Cambridge,
Madingley Rd, Cambridge, CB3 0HA, UK}
\altaffiltext{3}{CERN, 1211 Gen{\`e}ve, Switzerland}
\altaffiltext{4}{Astronomy Unit, School of Mathematical Sciences, Queen
Mary, University of London, Mile End Road, London E1 4NS, UK}
\altaffiltext{5}{Laboratoire de Physique Corpusculaire et Cosmologie,
Coll{\`e}ge de France, 11 Place Marcelin Berthelot, F-75231 Paris,
France}
\altaffiltext{6}{Department of Astronomy, Ohio State University, 140
West 18th Avenue, Columbus, OH 43210}
\altaffiltext{7}{Laboratoire d'Astrophysique UMR~CNRS~5572,
Observatoire Midi-Pyr{\'e}n{\'e}es, 14 Avenue Edouard Belin, F-31400
Toulouse, France}

\begin{abstract}
The POINT-AGAPE collaboration is currently searching for massive
compact halo objects (MACHOs) towards the Andromeda galaxy (M31). The
survey aims to exploit the high inclination of the M31 disk, which
causes an asymmetry in the spatial distribution of M31 MACHOs.  Here,
we investigate the effects of halo velocity anisotropy and flattening
on the asymmetry signal using simple halo models.  For a spherically
symmetric and isotropic halo, we find that the underlying
pixel-lensing rate in far-disk M31 MACHOs is more than 5 times the
rate of near-disk events. We find that the asymmetry is increased
further by about 30\% if the MACHOs occupy radial orbits rather than
tangential orbits, but is substantially reduced if the MACHOs lie in a
flattened halo. However, even for haloes with a minor-to-major axis
ratio $q = 0.3$, the numbers of M31 MACHOs in the far-side outnumber
those in the near-side by a factor of $\sim$2. There is also a
distance asymmetry, in that the events on the far-side are typically
further from the major axis.  We show that, if this positional
information is exploited in addition to number counts, then the number
of candidate events required to confirm asymmetry for a range of
flattened and anisotropic halo models is achievable, even with
significant contamination by variable stars and foreground
microlensing events.  For pixel-lensing surveys which probe a
representative portion of the M31 disk, a sample of around 50
candidates is likely to be sufficient to detect asymmetry within
spherical haloes, even if half the sample is contaminated, or to
detect asymmetry in haloes as flat as $q = 0.3$ provided less than a third of
the sample comprises contaminants. We also argue that, provided its
mass-to-light ratio is less than 100, the recently observed stellar
stream around M31 is not problematic for the detection of asymmetry.
\end{abstract}

\keywords{
Dark Matter -- Galaxies: Individual (M31) -- Gravitational Lensing}


\section{Introduction}

Continuing disagreement as to whether Massive Compact Halo Objects
(MACHOs) have been detected by microlensing experiments looking
towards the Magellanic Clouds highlights the need for other
microlensing targets (e.g., Kerins 2001). The Andromeda Galaxy (M31)
presents an opportune target in this respect. The disk of M31 is
highly inclined ($i\sim 77\degr$), with the consequence that lines of
sight to disk stars in the north-west or near side of M31 are shorter
than those to the south-east or far side. Microlensing by a spheroidal
dark halo will have a characteristic signature with an excess of
events on the far side of the M31 disk (Crotts 1992; Baillon et al.\
1993). This asymmetric signal is absent for variable stars or stellar
microlenses in the disk of M31. A number of groups (e.g., Auri\`ere et
al.\ 2001; Riffeser et al.\ 2001; Calchi-Novati et al.\ 2002; Crotts
et al.\ 2001) are now carrying out large-scale surveys of M31 to
search for this near-far disk asymmetry. This is a mammoth task as the
individual stars in M31 are not resolved, so that new
techniques based on the super-pixel method (Ansari et al.\ 1997) or
difference imaging (e.g., Crotts \& Tomaney 1996) have been exploited
to measure the flux changes on unresolved stars. Nonetheless,
convincing candidate events are now being discovered, for example by
the POINT-AGAPE collaboration (e.g., Auri\`ere et al.\ 2001;
Paulin-Henriksson et al.\ 2002, 2003). Therefore, this is a timely
moment to consider what factors affect the near-far disk asymmetry and how
many events are likely to be needed for a convincing detection.

The aim of this paper is to estimate the size of candidate event
samples needed to detect asymmetry for different halo models.  In
Sections 2 and 3 of the paper we show how the magnitude of the
asymmetry signal is affected by the velocity anisotropy and the
flattening of the M31 baryonic dark halo respectively. There are few
ways known to us for measuring the properties of the orbits of dark
objects in the halo of any galaxy or for ascertaining the flattening
of the baryonic dark component of the halo. Hence, any clues gleaned
from pixel lensing experiments will be invaluable. In Section 4, we
present some simple, non-parametric statistical estimators of
asymmetry and calculate how many candidates are needed to give convincing
detections.

\section{The Effects of Velocity Anisotropy} \label{pixelrate}

\subsection{Anisotropic Models}

Here, we investigate the microlensing properties of haloes in which
the velocity distribution of the MACHOs is anisotropic. We use models
in which the halo density $\rho$ is isothermal and the rotation curve
is flat:
\begin{equation}
\rho \propto \frac{v_0^2}{r^{2}}, \ \ \ v_0^2 = {\rm constant}.
\label{halo}
\end{equation}
For these models, the velocity dispersions in the spherical polar
coordinate system are given by (see e.g., White 1981; Evans, H\"afner
\& de Zeeuw 1997)
\begin{equation}
\sigma_\phi^2 = \sigma_\theta^2 = (1+\alpha)\,\sigma_r^2 =
\frac{v_0^2}{2},
\label{vel}
\end{equation}
where $\alpha > -1$ is the anisotropy parameter. If $\alpha < 0$, then
the velocity distribution is referred to as `radially anisotropic'; if
$\alpha > 0$, then it is `tangentially anisotropic.' Whilst
$\sigma_r^2$ in equation~(\ref{vel}) diverges as $\alpha \rightarrow
-1$, velocity dispersion ratios are rarely observed to be more extreme
than 3:1, implying $-8/9 < \alpha < 8$. We assume a circular velocity
of $v_0 = 235\ \mbox{km s$^{-1}$}$ for M31's halo (Emerson 1976) and
we compute the cutoff radius to give a total halo mass of $1 \times
10^{12}\ \sm$ (e.g., Evans \& Wilkinson 2000). The sources
are drawn from the M31 disk which is adequately modelled as a sheet
inclined at 77\degr\ to the line of sight. The source velocity is
assumed to be dominated by the disk rotation speed of 235~$\mbox{km
s$^{-1}$}$.

For each halo model, we calculate a theoretical estimate of the
pixel-lensing rate $\Gamma_{\rm p}$. Unlike the classical (resolved
star) microlensing rate, $\Gamma_{\rm p}$ depends additionally on the
surface brightness of the M31 disk and the luminosity function of the
M31 sources (c.f., Kerins et al.\ 2001). The calculations here are
performed for a $V$-band luminosity function and surface brightness
distribution. The surface brightness of M31 is tabulated in Walterbos
\& Kennicutt (1987). The M31 disk luminosity function is assumed to be
the same as for the Milky Way and we use the data of Wielen, Jahrei\ss\
\& Kruger (1983) to characterise the faint end ($M_V > 5$) and that of
Bahcall \& Soneira (1980) for the bright end. For each source star of
flux $F$ at sky position ($x$,$y$) on the M31 disk, we compute the
maximum impact parameter, $u_{\rm T}$, needed to ensure that the
magnified source star noticeably enhances the local flux contribution
from the background galaxy and sky background. Motivated by the
POINT-AGAPE experiment (e.g., Paulin-Henriksson et al.\ 2002, 2003),
we assume that detection is performed on a ``super-pixel'' array of
pixels of size $2\farcs1 \times 2\farcs1$ which typically encloses
40\% of the total flux from a point source. In addition to the
background due to the M31 surface brightness we allow for a sky
background of 19.5~$\mbox{mag arcsec$^{-2}$}$. We assume an event is
detectable if the flux change caused by microlensing is $\ga 1\%$ of
the background flux on the super-pixel. The pixel lensing rate is then
\begin{equation}
\Gamma_{\rm p}(x,y) =
\langle u_{\rm T}(x,y) \rangle \Gamma_{\rm c}(x,y),\label{prate}
\end{equation}
where $\Gamma_{\rm c}$ is the classical microlensing rate (Paczy\'nski
1986; Kiraga \& Paczy\'nski 1994) and $\langle u_{\rm T} \rangle$ is
the maximum impact parameter averaged over the luminosity function
$\phi$ of the source stars. Explicitly, we can write that
\begin{equation}
\langle u_{\rm T}(x,y) \rangle =
{\int \phi(F)\,u_{\rm T}(F,x,y)\,dF \over \int \phi(F)\,dF}.
\end{equation}
This is an upper limit to the observed pixel lensing rate for any real
experiment, as it does not take into account the effects of sampling,
changing observing conditions or event identification algorithms.
These effects may alter the spatial distribution of events, but they
can be corrected via the calculated detection efficiency. In fact, the
efficiency is largely controlled by the local surface brightness and
so is approximately symmetric with respect to the major axis. The
ratio of the number of far-disk to near-disk events therefore does not
depend on the efficiency to lowest order.

Finally, the spatially-averaged pixel-lensing rate $\pixrate$
is obtained by weighting the rate with the M31
disk surface brightness. The central portions of the M31 disk are
omitted, partly because stellar lenses in the M31 bulge dominate here
and partly because the halo model is singular at the centre. So, a
central region of 5\arcmin\ radius is excised from the M31 disk before
performing the spatial averaging separately for events above and below
the M31 major axis. None of the current experiments is surveying the
entire M31 disk, but their fields do span the large majority of the
minor axis and we therefore expect any underlying asymmetry in their
fields to be representative of the globally-averaged asymmetry
computed here.

\subsection{Results}

\begin{figure}
\begin{center}
\includegraphics[scale=0.5,angle=270]{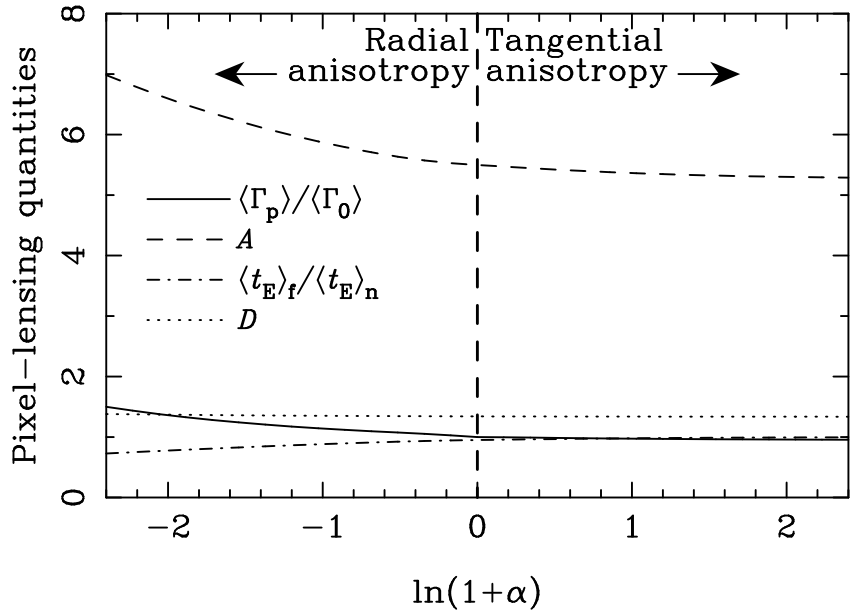}
\caption{The spatially-averaged pixel-lensing rate $\pixrate$
({\it solid line}) as a function of anisotropy
parameter $\alpha$ and normalised to the value for the isotropic model
$\langle \Gamma_0 \rangle = 
7.6 \times 10^{-7}\ \mbox{stars$^{-1}$ year$^{-1}$}$.
Also shown is the M31 MACHO number asymmetry $A$ ({\it dashed line}), 
the ratio of near-disk to far-disk average durations $\timeratio$ ({\it dot-dashed line}), and the ratio of projected distances to the major axis $D$ ({\it dotted line})}.  

\label{fig:pix}
\end{center}
\end{figure}

Figure~\ref{fig:pix} shows the spatially-averaged theoretical
pixel-lensing rate, $\pixrate$, for lenses of mass $M$, normalised to
the value $\Gamma_0 = 7.6 \times 10^{-7}\ (M/\sm)^{-1/2}$ events per
star per year. When $\alpha < 0$, the velocity distribution is
radially anisotropic and the rate $\pixrate \propto v_0 /
\sqrt{1+\alpha}$. When $\alpha >0$, the velocity distribution is
tangentially anisotropic and the rate $\pixrate \propto v_0$, so the total rate in radially anisotropic models is
somewhat higher.  However, we are primarily interested in the
differences between the near and far disk. Such asymmetries may
manifest themselves in the numbers, locations and time-scales
of the events. Accordingly, Figure~\ref{fig:pix} also shows the far-to-near disk ratios  for the pixel-lensing rates ($A =
\pixratef / \pixraten$), the mean Einstein crossing times
($\timeratio$) and the ratio of mean projected distances of events to the major
axis ($D$).

All models show a strong excess of far-disk events, with $A$
increasing from 5.3 to 7 as the models go from tangential to radial
anisotropy. It is also evident from $D$ in Figure~\ref{fig:pix} that
far-disk events lie systematically farther from the major axis than
near-disk events, providing a second signature of asymmetry. However
$D$ is nearly constant across the range of $\alpha$ so this spatial signature
does not provide a probe of the degree of velocity anisotropy.

There is also an asymmetry in the Einstein crossing times of M31
MACHOs. If MACHOs have radially distended orbits, then their motion
tends towards being parallel to the line of sight for the near disk but
orthogonal to it for the far disk. Events on the near side therefore last
longer. In other words, the ratio of the typical time-scales of
far-disk to near-disk events decreases with increasing radial
anisotropy.  Consequently, the ratio of the numbers of events in the
far-disk to the near-disk is enhanced in radially anisotropic models
compared to isotropic model. This can be seen from the fact that the
spatially-averaged microlensing optical depth, $\langle \tau \rangle
\propto \langle \te \rangle \pixrate$, is independent of the velocity
distribution. From Figure~\ref{fig:pix}, the expected enhancement in
the number asymmetry $A$ is $\sim$30\%.

For the isotropic model ($\alpha = 0$), the time-scales are
typically shorter in the far disk than in the near disk. The
reason for this is that the typical separation
between lens and source is larger for near-disk MACHOs. For far-disk
events, the typical lens-source separation is biased towards the
location where the density peaks along the line of sight, which is at
a distance $\sim |y| \tan i$ in front of the sources, where $y$ is the
projected distance along the minor axis. The situation is different
for near-disk MACHOs, where the line of sight density is always a
monotonically decreasing function of lens-source separation. Here, the
typical separation is $\sim$20--30~kpc for $y > 5\arcmin$, so $\timeratio$ is less than unity for an isotropic model,
as shown in Figure~\ref{fig:pix}.

Unfortunately, the Einstein crossing time is not generally measurable
for pixel-lensing events. Instead experiments measure the full-width
half-maximum ($t_{1/2}$) of the lightcurve, which is additionally
correlated with the source luminosity and background surface
brightness distribution. The resulting $t_{1/2}$ distributions are
therefore predicted to be broad and any asymmetries
in the time-scale distribution are unlikely to be easily observable.

\section{The Effects of Flattening}

\subsection{Flattened Models} \label{tilt}

Self-consistent solutions of the self-gravitation equations for the
density, potential and velocity distributions of flattened halo models
are rare. Even solving the Jeans equations for the second velocity
moments can lead to cumbersome results. Baltz, Gyuk \& Crotts (2003)
have computed pixel-lensing rates explicitly for simple analytic
flattened halo models. However, if we are merely interested in
comparing the effects of flattening on the ratios of quantities in the
near- and far-disk, then there is a quick alternative to carrying out
computations with a fully axisymmetric halo model.

\begin{figure}
\begin{center}
\includegraphics[scale=0.6,angle=0]{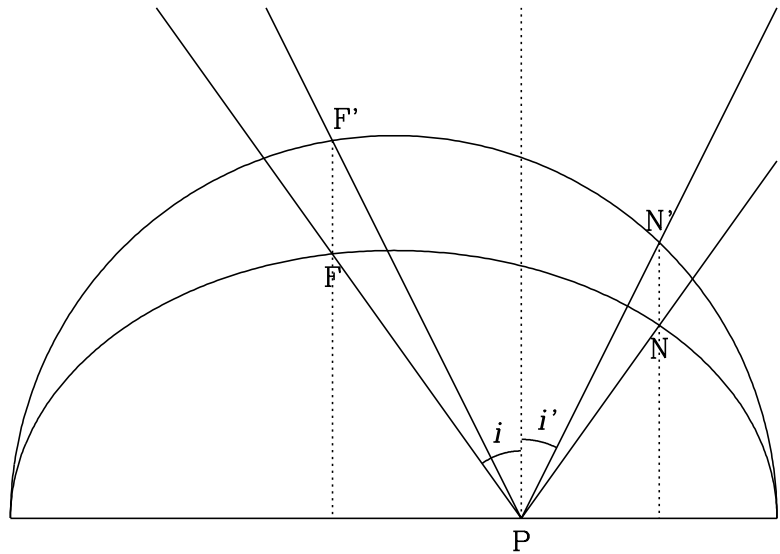}
\caption{This shows lines of sight through a spherical and an
elliptical halo. From the properties of an ellipse, we know that the
ratios NP:FP and N'P:F'P are equal. This enables us to relate the
asymmetry signal of a disk viewed through a flattened halo at
inclination $i$ to the same disk viewed through a spherical halo at
inclination $i'$.}
\label{fig:schematic}
\end{center}
\end{figure}

Figure~\ref{fig:schematic} shows lines of sight passing through an
elliptical halo with axis ratio $q$ and eccentricity $(1-q^2)^{1/2}$.
The lines strike the disk of M31 at an angle $i$. The ratios of the
optical depths in the near and far disk are proportional to the line
segments NP and FP. Also shown is an equivalent spherical halo. By
extending N vertically to N' and F to F', we can construct lines of
sight that pass through the spherical halo and strike the disk at a
different angle $i'$. Similarly, the ratios of the optical depths are
proportional to the line segments N'P and F'P. From the elementary
properties of the ellipse, it follows that the two ratios are the same
(NP/FP = N'P / F'P). By straightforward trigonometry, one has $q \tan i = \tan i'$. In other words, the asymmetry signal of a
flattened halo with axis ratio $q$ is the same as that of an
equivalent spherical halo, provided the disk is viewed not at angle
$i$ but at an angle $i'$.

This transformation takes into account the geometric effects of
flattening. The first-order changes in the velocity distribution can
be computed using the tensor virial theorem (Binney \& Tremaine 1987;
Han \& Gould 1995). Strictly speaking, the tensor virial theorem
applies globally and relates the components of the total kinetic
energy tensor $T$ to the components of the total potential energy
tensor $W$. If we assume that the virial theorem holds at each spot,
then it follows that
\begin{equation}
{\sigma_x^2 \over \sigma_z^2} \approx {T_{xx} \over T_{zz}} = {W_{xx}
\over W_{zz}} \approx {1 \over q^2}.
\end{equation}
Here, we have assumed that the figure is oblate spheroidal, with the
the short axis being in the $z$ direction and the $(x,y)$ plane being
equatorial. Although this is only an approximate relation, valid
for small flattening, it shows that the first-order changes in the
velocity distribution are also accounted for by the transformation.

This means that a quick way to study the asymmetry properties of flattened haloes
is to take isotropic spherical models and vary the inclination angle
of the M31 disk. We stress that the transformation does not allow us
to calculate absolute quantities like the rate, but only the
ratios of such quantities in the near and far disk.

\subsection{Results}

\begin{figure}
\begin{center}
\includegraphics[scale=0.5,angle=270]{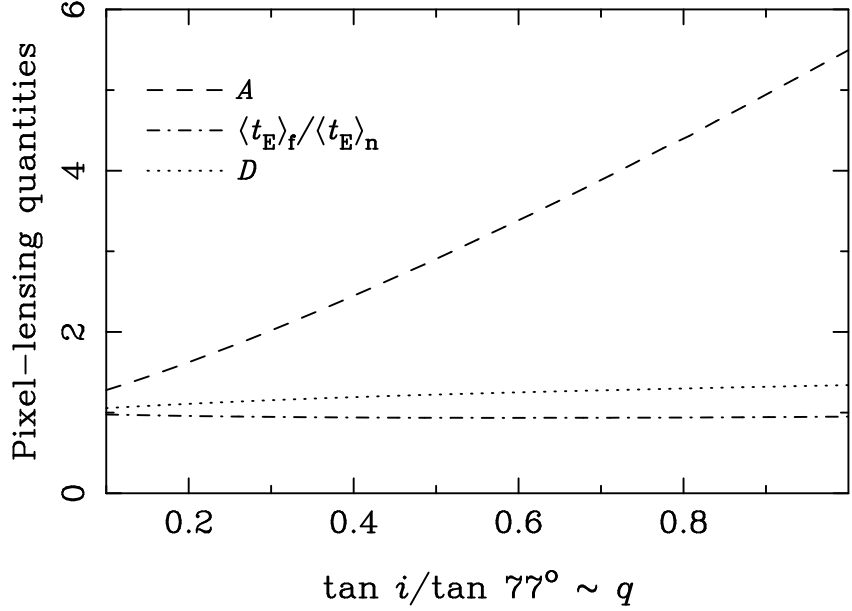}
\caption{The far-to-near-disk ratio for the pixel lensing rate $A$ ({\it dashed line}), the
mean vertical distance $D$ ({\it dotted line}) and the time-scales
({\it dot-dashed line}),
shown as a function of flattening. This diagram is drawn using the
transformation introduced in Section~\ref{tilt}.}
\label{fig:flatpix}
\end{center}
\end{figure}

Figure~\ref{fig:flatpix} shows the variation in the far-to-near-disk ratios for the pixel
lensing rates $A$, the means of vertical distances $D$ and the
time-scales.  The asymmetry signal
$A$ for a halo with flattening $q$ is linearly related to the
asymmetry signal for a spherical halo $A_0$, that is
\begin{equation}
A \simeq 1 + q(A_0-1).
\end{equation}
The asymmetry signal clearly diminishes with flattening; it is obvious
that, in the completely flat limit where the halo becomes a razor-thin
disk, the asymmetry must vanish. The change in $A$ with flattening is
almost entirely caused by the change in the ratio of optical
depths. Therefore the ratio of the average time-scales is largely unaffected
by flattening.

The distance asymmetry $D$ also decreases with increasing flattening,
as is again obvious in the razor-thin limit. The distance asymmetry
arises in a spherical model because of two effects. First, lines of
sight are longer as we move from near to far disk. Second, the line of
sight with greatest column density goes through the centre in the
near-disk but lies away from the centre in the far-disk. So the
distribution of distances of events is monotonically decreasing in the
near disk, but rises to a maximum and then decreases in the far
disk. As the flattening increases, all lines of sights become shorter
and the density becomes more concentrated towards the centre. The
distributions of distances in both the near and far disk shrink and
the maximum moves towards the centre in the far disk. This latter
effect is the dominant one, and so the distance asymmetry signals
falls with increasing flattening.

\section{Signal Detection}

\subsection{The Number Asymmetry Signal}

In this section, we ask whether one can detect the asymmetry signals
in the presence of contaminating events and how the velocity
anisotropy and the halo flattening affect this detectability. That is,
we ask how many candidates are needed to detect the
asymmetry at a certain confidence level.

First, we consider the number asymmetry $A$ between the far and near
side. One problem is that none of the experiments will be able to
obtain pure samples of M31 MACHO events; some microlensing
contamination from Milky Way MACHOs and M31 stars, as well as
non-microlensing contamination from variable stars and supernovae in
background galaxies will be inevitable. Contamination by periodic
variables can be minimized by observing over a sufficiently long
time. Colour information can also be used to eliminate variable stars.
Fortunately, the other contaminants are equally likely to occur in the
near and far regions of the M31 disk. Whether or not an asymmetry can
be detected is therefore a question of the size of the sample, the
magnitude of the underlying M31 MACHO asymmetry and the level of
contamination.

\begin{figure}
\begin{center}
\includegraphics[scale=0.5,angle=270]{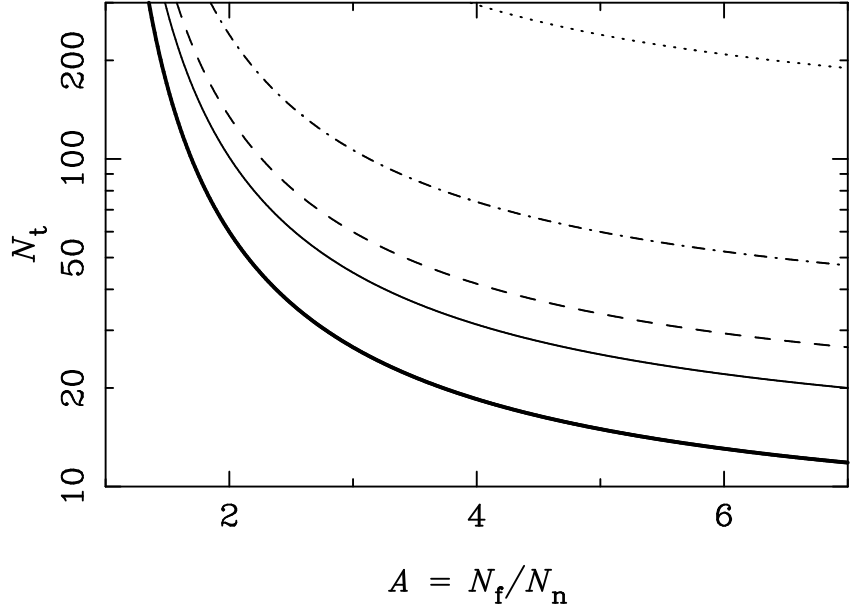}
\caption{The number of candidate events (M31 MACHOs and
contaminants) required to confirm asymmetry with 99\% confidence,
based upon number-count information alone, plotted as a function of the underlying M31 MACHO number
asymmetry $A$. The different lines
correspond to contamination factors, $\fc = 0$ ({\it thick solid
line}), 0.3 ({\it thin solid line}), 0.5 ({\it dashed line}), 1 ({\it
dot-dashed line}) and 3 ({\it dotted line}).}
\label{fig:crit}
\end{center}
\end{figure}

Let us denote the numbers of M31 MACHO events in the near and far disk
by $\Nn$ and $\Nf$ respectively, and the number of contaminants by
$\Nc$. The condition for a detection of the asymmetry signal at the
$s$-$\sigma$ level is that the difference between the far and the near
counts be greater than $s$ times the Poisson error, which is given by
the square root of the total number of the events. Therefore
\begin{equation}
{\Nf^\prime - \Nn^\prime \over \sqrt{\Nf + \Nn + \Nc}} > s
\label{asymm_c}
\end{equation}
where $\Nf'$ and $\Nn'$ are the total number of
candidates on the far and near sides, respectively, including
contaminant events.
In the case where the contaminants are distributed evenly between the
far and near disk, $\Nf' - \Nn' = \Nf - \Nn$. Straightforward
manipulation of the above condition leads to a condition on the
total number of candidate events,
\begin{equation}
\Nt > s^2 \left({A+1 \over A-1}\right)^2 (1 + \fc)^2,
\label{asymm}
\end{equation}
where $A = \Nf/\Nn$ is the underlying asymmetry of the M31 MACHO
events, $\fc = \Nc/(\Nf +\Nn)$ is the contamination factor, and $\Nt =
\Nf+\Nn+\Nc$ the total number of candidate events.
Figure~\ref{fig:crit} shows the expected number of events
(including the contaminants) required to detect an asymmetry at the
99\% confidence level ($s=2.58$) for various contamination factors. We
see immediately that the size of the asymmetry signal is crucial. If
$A \la 2$, then even with low contamination the number of M31
microlensing events needed to give a convincing detection of asymmetry
exceeds $60$. On the other hand, if the asymmetry is $\sim$5, then more than 15 candidates are required. Reassuringly, on referring to Figures~\ref{fig:pix} and
\ref{fig:flatpix}, we see that $A$ typically lies between 5 and 7 for
spherical models, and only becomes as small as 2 for models with
$q = 0.3$. In fact, haloes flatter than this
are not likely on dynamical grounds, as they are susceptible to
bending instabilities (e.g., Merritt \& Sellwood 1994). Samples of
$\sim$15 candidate events are well within reach of the current surveys
if MACHOs contribute significantly to the dark matter budget. However,
this number is only the contribution of the M31 halo, as we have cut
out the central parts of the M31 disk.

If the contribution of MACHOs to the dark matter mass budget is
significant in M31, we expect a similar MACHO contribution for the
Milky Way. For interesting MACHO fractions, the foreground Milky Way
MACHOs may well provide the dominant contribution to $\fc$. If the
typical MACHO mass and halo density contributions are universal, then the magnitude of $\fc$ will be determined, to first
order, by the relative masses of the M31 and Milky Way
haloes. If the M31 halo is twice as massive as the Milky Way's, then
$\fc \sim 0.4$ (Kerins et al.\ 2001), whilst we should expect $\fc$ to
be closer to unity if the two haloes are
equally massive (Evans \& Wilkinson 2000). In any case, the total number of candidates
required to confirm asymmetry scales as $(1+\fc)^2$ from
equation~(\ref{asymm}), so we require four times as many candidates to
detect the asymmetry when $\fc = 1$ as when $\fc = 0$.

\subsection{The Distance Asymmetry Signal}

Motivated by the variation in mean distances of events between the
near and far disk seen in Figures~\ref{fig:pix} and
\ref{fig:flatpix}, we can go beyond the asymmetry in number counts by
considering a distance asymmetry signature.

To quantify differences between the near-disk and far-disk event
positions, we apply the Mann-Whitney rank-sum test (Mann \& Whitney
1947). This is a non-parametric test for differences between the medians of two
samples. The Mann-Whitney test exploits the fact that two samples
drawn from identical distributions exhibit the property that, if one
combines them and then ranks the elements by size, the two samples are
uniformly intermingled on average within the ranked combined
sample. For two samples of large enough size $n_a$ and $n_b$, the sum
of the rank numbers for sample $a$ is normally distributed about a
mean $\mu_a = n_a (n_a + n_b)/2$ with a variance $\sigma_a^2 = n_a n_b
(n_a+n_b+1)/12$. The null hypothesis of similarity (or, in our case,
symmetry) is therefore straightforward to quantify.

In the case of M31 pixel lensing, we can apply the Mann-Whitney test
to the distribution of $y$, the projected distance from the M31 major
axis, for candidates in the near- and far-disk sub-samples.  We can
also combine the Mann-Whitney test with the number asymmetry test, since the Mann-Whitney statistic probes the spatial
distribution of events, not the relative sizes of the samples.  If the
far-disk event positions are designated as sample $a$ and the sum of
their rank numbers within the combined near- and far-disk sample is
$\theta_a$, then we can define $s_{\rm MW} = (\theta_a -
\mu_a)/\sigma_a$. Taking $s_{\rm N}$ to be the significance of the
number asymmetry statistic, as defined by the left-hand side of
equation~(\ref{asymm_c}), then the overall significance of the
combined sample is $s = (s_{\rm MW}^2 +s_{\rm N}^2)^{1/2}$.  A value
of $s = 2.58$ would indicate a near-far asymmetry favoured at the 99\%
confidence level.

We have performed Monte Carlo simulations to determine the number of
candidates required to secure a 99\% confidence detection of
asymmetry. The simulations test a range of anisotropic and flattened
halo models and a range of contaminations. For a given model, event
position realisations are generated using the theoretical
pixel-lensing rate of equation~(\ref{prate}), weighted by source
number density. For each new event realisation, the overall
significance, $s$, of the cumulative sample is computed from both the
Mann-Whitney statistic ($s_{\rm MW}$) and the number asymmetry
statistic ($s_{\rm N}$). When there are fewer than five events in
either the far- or near-disk sub-samples, only the number asymmetry
statistic is used because the distribution of rank sums for small
data-sets can deviate strongly from Gaussianity.  We assume the
contaminating populations are symmetrically distributed about the M31
major axis, though we adopt the most difficult case when they have a
comparable spatial dispersion to the M31 MACHO events. The extent to
which this is true depends upon whether Milky Way MACHOs, on the one
hand, or variable stars and stellar microlenses, on the other hand, provide
the dominant contribution to $\Nc$. If the latter population dominates,
then contaminants are likely to be more spatially concentrated than
M31 MACHOs, making the MACHO asymmetry easier to measure for a given
$\fc$. So for each pixel-lensing realisation, a uniform random number
in the interval [0,1] is chosen. If it is less than $\fc/[2(1+\fc)]$,
then the position of the event is flipped about the M31 major
axis. This means that, on average, a fraction $\fc/(1+\fc)$ of the
total sample is symmetrically distributed, as required. A full trial,
$i$, is terminated either when the cumulative sample of $N_{{\rm
t},i}$ candidates provides at least a 99\% confidence detection of
asymmetry, or when the program estimates that the required sample is
likely to exceed 500 events. This whole process is repeated 1000 times
for each model and the median value of $N_{{\rm t},i}$ (excluding
trials which are prematurely terminated) is adopted as the estimate
for $\Nt$.

\begin{figure}
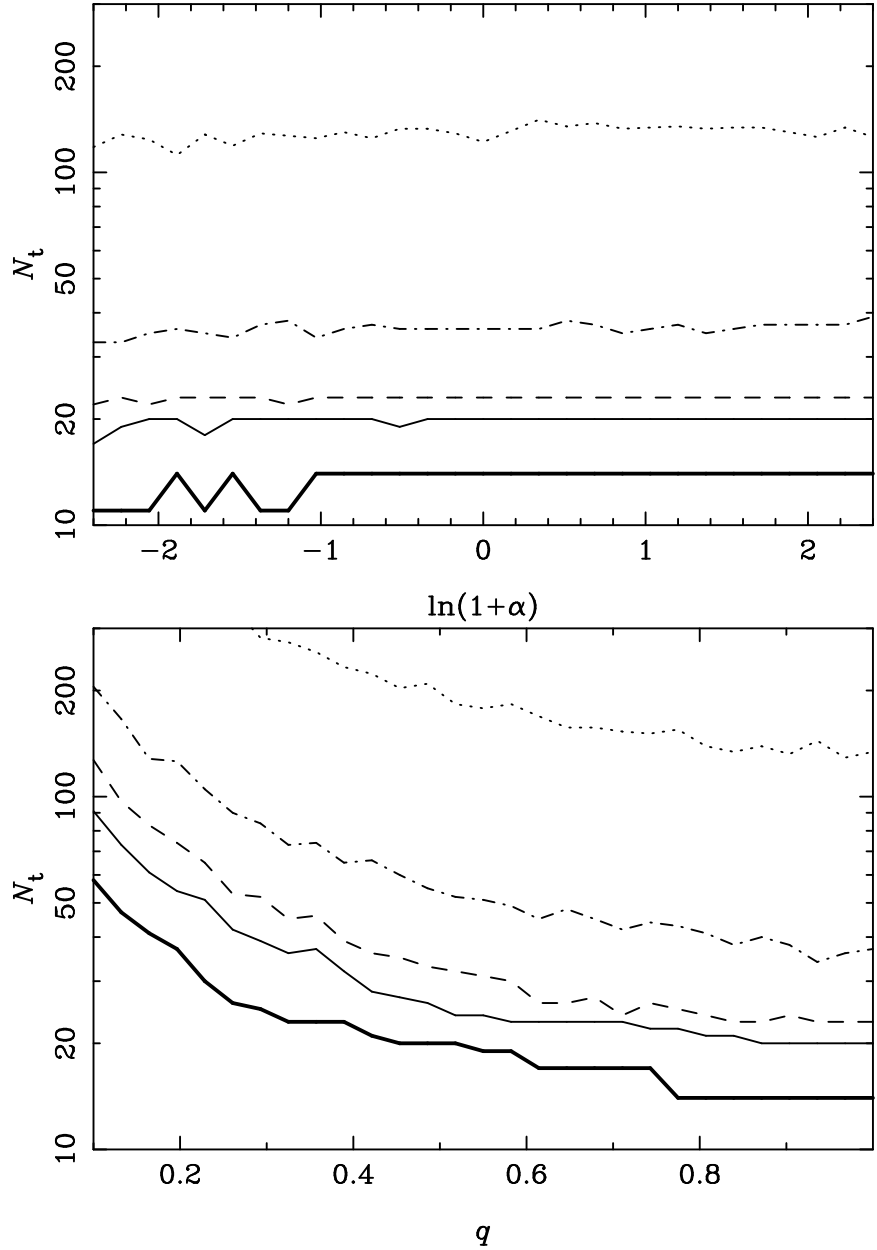

\begin{center}
\includegraphics[scale=0.5,angle=270]{mw-anisop.ps}\hspace{1.cm}
\includegraphics[scale=0.5,angle=270]{mw-flat.ps}
\caption{The median number of candidates (M31 MACHOs and
contaminants) required for a 99\% confidence detection of
asymmetry. Line coding is the same as in Figure~\ref{fig:crit}. {\it
Upper panel}\/ shows $\Nt$ as a function of
halo velocity anisotropy for a spherical halo. {\it Lower panel}\/
shows $\Nt$ for flattened halo models, using the equivalence
between the halo flattening parameter $q$ and disk inclination $i$
discussed in Section~\ref{tilt}.}
\label{fig:monte}
\end{center}
\end{figure}

The upper panel of Figure~\ref{fig:monte} shows the median number of
candidate events (including the contaminants) required to detect an
asymmetry with 99\% confidence, as a function of the halo velocity
anisotropy parameter $\alpha$. In the absence of contamination, a
sample size between 11 to 14 events is typically sufficient to detect
asymmetry.  The raggedness of this line is partly due to Monte Carlo
noise but also partly due to the fact that $\Nt$ necessarily takes
only a few discrete values in the limit of small data-sets. From
equation~(\ref{asymm_c}), the smallest sample needed give rise to a
99\% confidence detection of asymmetry is 7 events, all of which must
be located in the far disk (assuming the asymmetry is caused by M31
MACHOs). When one candidate lies in the near disk we need $\Nt = 11$,
and when two candidates lie in the near disk we require $\Nt =
14$. The leaps in the thick solid line between 11 and 14 events
reflect this discrete behaviour, though the oscillation back and forth
for $-2< \ln (1+\alpha) < -1$ is due to Monte-Carlo noise.  In the
worst case considered in Figure~\ref{fig:monte}, where contaminants
outnumber M31 MACHOs by 3:1, a median sample of $\sim$120 events is
required to detect asymmetry.  For $\alpha = 0$ and $\fc = 3$, we find
that the expectation value of $\Nt$ is around 190. Using $s = 2.58$,
$\fc = 3$ and an asymmetry $A = 5.5$ when $\alpha = 0$
(Figure~\ref{fig:pix}), equation~(\ref{asymm}) indicates that we
should require $\Nt = 220$ if we use number-count information alone.
Therefore the addition of distance information allows around a 15\%
reduction in the required number of candidates in this case.

Overall, $\Nt$ does not appear to be particularly sensitive to
$\alpha$. This is to be expected when contamination levels are high
because the 30\% contrast in $A$ between the radially- and
circularly-anisotropic models (see Figure~\ref{fig:pix}) is strongly
diluted. 

The lower panel of Figure~\ref{fig:monte} shows the situation for
flattened halo models. The larger range in $A$ for the flattened
models ($1.3 < A < 5.5$) means that the median $\Nt$ shows a greater
sensitivity to flattening than to velocity anisotropy. In the absence
of contaminants, $\Nt \la 25$ is typically
needed to confirm asymmetry for models rounder than $q = 0.3$. A sample of
100 candidates would permit asymmetry to be detected even if $\fc
\simeq 1$. Adding distance information is particularly effective at
reducing the required size of candidate samples for highly flattened
models. As an extreme example, when $q = 0.1$ and $\fc = 0$, the
expectation of $\Nt$ is around 80 if distance information
is used along with number counts. From equation~(\ref{asymm}), 750
candidates are required if only number counts are used. This shows the
value of the Mann-Whitney statistic.

It is of course much more difficult to measure an asymmetry than
to detect one.  Suppose an experiment has gathered
$\Nt \ga 100$ events (including contaminants). With reference to
Figure~\ref{fig:crit}, if there has been no detection of asymmetry at the 99\%
confidence level, then -- as $A \ga 3$ for all models we have considered
with $q > 0.5$ -- we can infer that the signal has been overwhelmed by
contaminants ($\fc > 1$). The degree of contamination must be greater
than a critical value which is given by computing the curve which
passes through the point ($A =3, \Nt$). So, a null signal can be used
to give a constraint on the contamination. On the other hand, if there
has been a detection, then the measured signal $A' \simeq (\Nf + \half
\Nc)/ (\Nn + \half \Nc)$ is merely a lower limit to the true signal.
It should be possible to estimate the contamination fraction
statistically even if we do not know the individual contaminating
events, so the true asymmetry signal can be matched to models using
standard Bayesian likelihood methods (Kerins et al 2001). Current
surveys should therefore be able to discriminate between halo models
with different degrees of flattening and may, if the spatial
distribution of contaminant populations is well characterized, be able
to distinguish between radially-anisotropic halo models and
tangentially-anisotropic or isotropic models.

\subsection{Confusion from M32 and its Stream}

Additional confusion of the asymmetry signal may come from
microlensing by stars which belong to streams or tidal debris from
disrupted satellite galaxies cannibalized by M31 and/or by stars
belonging to the intervening dwarf elliptical M32.

Ibata et al.\ (2001) have traced out a giant stream in the M31 stellar
halo in red giant branch star counts. The stream is $\sim$1\degr\ wide
in projection. It seemingly originates from the satellite galaxy M32
and possibly also incorporates NGC~205. The average surface brightness
of the stream is $\sim$30~$\mbox{mag arcsec$^{-2}$}$  in the
$V$-band. It has been suggested by Ferguson et al.\ (2002) that this
may confuse the detection of the near-far disk asymmetry in the
microlensing experiments. For sources in M31 and lensing populations
at roughly the same distance from the sources, the optical depth of
the stream is
\begin{equation}
\tau \sim 4.3 \times 10^{-10} \left( {d \over 20\ \mbox{kpc} }\right)
\left ( {{\rm M/L} \over \sm / \mbox{L}_{\odot}} \right) 10^{12 -0.4
\mu}
\end{equation}
where $\mu$ is the surface brightness in magnitude per arcsec$^2$,
$d$ the separation of the stream from M31 disk along the line of sight
and M/L is the MACHO mass-to- light ratio of the stream. This
ratio must exceed 100 for the optical depth of the stream to be
comparable to the other lensing populations. Accordingly, the presence
of such streams is not likely to be problematic for asymmetry
detection.

More problematic may be the intervening dwarf elliptical galaxy
M32. The POINT-AGAPE collaboration has already found
a candidate which lies $\sim$3\arcmin\ in projection away from
the centre of M32 and argued that the lens most probably lies in M32
itself (Paulin-Henrikkson et al.\ 2002). The optical depth of M32 is
estimated to be (Paulin-Henrikkson et al.\ 2003)
\begin{equation}
\tau \sim 1.4 \times 10^{-6} \left( {d \over 20\ \mbox{kpc} }\right)
\left ( {{\rm M/L} \over 3\sm / \mbox{L}_{\odot} }\right)
\end{equation}
where $d$ is the separation of M32 from M31 along the line of sight
and $M/L$ is the stellar mass-to-light ratio of M32. This is comparable to the signal expected from the
baryonic halo (assuming a 20\% MACHO fraction). 
Although microlensing associated with M32 is a potentially significant 
contaminent the affected region can be readily excluded from the statistics
using an appropriate mask. 

\section{Conclusions}

Pixel-lensing experiments targeting M31 are
hoping to exploit the favourably high disk inclination in order to
detect an asymmetry in the spatial distribution of microlensing
events. If such a signal is found, it will provide powerful evidence
for the existence of MACHOs.

For a spherically symmetric and isotropic halo, the numbers of M31
MACHOs in the far-disk outnumber those in the near-disk by more than 5
to 1. This asymmetry is increased by about 30\% if M31 MACHOs occupy
radial orbits rather than tangential ones. The signal is diminished if M31
MACHOs lie in a flattened halo. However, even for haloes as flat as
$q = 0.3$, the numbers of M31 MACHOs in the far side will outnumber those in
the near side by a factor of $\sim$2.

The key to detecting asymmetry is to isolate microlensing events
solely due to M31 MACHOs. There is likely to be significant
contamination from other microlensing populations, as well as from
variable stars and supernovae mistaken for microlensing, which will
dilute the observed signal. The combination of number-count and
distance information permits asymmetry to be detected for a wide range
of halo models, even in the presence of significant levels of
contamination. For models with high levels of asymmetry, such as
spherical haloes or haloes with a high degree of radial velocity
anisotropy, number count information alone provides a sensitive
diagnostic. The addition of distance information allows $\sim$15\%
reduction in the size of samples needed to confirm asymmetry. For
models with low levels of asymmetry, such as strongly flattened
haloes, distance information can reduce the required size of candidate
samples by a factor 2 or more.

A sample of 50 events is typically sufficient to detect asymmetry in
the M31 MACHO distribution within spherical haloes, even if only half
the sample is due to M31 MACHO events. For flattened halo models, a
sample of 50 candidates would likely allow asymmetry to be seen,
provided that the halo axis ratio $q \geq 0.3$ and the contaminants do
not contribute more than a third of the sample. The term
``contaminants'' covers Milky Way MACHOs, M31 disk stellar lensing
events and variable star populations, all of which are assumed to be
symmetrically distributed with respect to the major axis. For
comparison, Paulin-Henriksson et al.\ (2003) have already found 362
lightcurves compatible with microlensing from the first two years of
the POINT-AGAPE survey, though the contamination factor may still be
very large.  Samples of 50 events with modest contamination are easily
achievable with the current generation of pixel-lensing surveys.

\acknowledgements NWE and JA thank the Royal Society and the
Leverhulme Foundation respectively for financial support.


\clearpage



\begin{thebibliography}{}

\bibitem[Ansari et al.(1997)]{ansari}
Ansari R. et al., 1997, A\&A, 324, 843

\bibitem[Auri\`ere et al.(2001)]{ma}
Auri\`ere M. et al., 2001, ApJ, 553, L137

\bibitem[Bahcall \& Soneira(1980)]{bahcall}
Bahcall J. N., Soneira R. M., 1980, ApJ, 238, L17

\bibitem[Baillon et al.(1993)]{baillon}
Baillon P., Bouquet A., Giraud-H\'eraud Y., Kaplan J., 1993, A\&A, 277, 1

\bibitem[Baltz, Gyuk \& Crotts(2003)]{baltz03}
Baltz E., Gyuk G., Crotts A., 2003, ApJ, 582, 30

\bibitem[Binney \& Tremaine(1987)]{galactic_dynamics}
Binney J., Tremaine S., 1987, Galactic Dynamics.
Princeton Univ.\ Press, Princeton, NJ

\bibitem[Calchi-Novati et al.(2002)]{sebastiano}
Calchi-Novati S. et al., 2002, A\&A, 381, 848

\bibitem[Crotts(1992)]{crotts}
Crotts A. P. S., 1992, ApJ, 399, L43

\bibitem[Crotts et al.(2001)]{crotts2}
Crotts A. P. S., 2001,
in Menzies J.W., Sackett P.D., eds., ASP Conf.\ Ser.\ Vol.\ 239,
Microlensing 2000: A New Era of Microlensing Astrophysics.
Astron.\ Soc.\ Pac., San Francisco, p.318

\bibitem[Crotts \& Tomaney(1996)]{crotts3}
Crotts A. P. S., Tomaney A. B., 1996, ApJ, 473, L87

\bibitem[Emerson(1976)]{1976MNRAS.176..321E}
Emerson, D. T., 1976, MNRAS, 176, 321 

\bibitem[Evans, H\"fner, \& de Zeeuw(1997)]{ehz}
Evans N. W., H\"afner R. M., de Zeeuw P. T., 1997, MNRAS, 286, 315

\bibitem[Evans \& Wilkinson(2000)]{evwil}
Evans N. W., Wilkinson M., 2000, MNRAS, 316, 929

\bibitem[Ferguson et al.(2002)]{2002AJ....124.1452F}
Ferguson A. M. N., Irwin M. J., Ibata R. A., Lewis G. F., Tanvir N. R.,
2002, AJ, 124, 1452

\bibitem[Han \& Gould(1995)]{hg}
Han C., Gould A., 1995, ApJ, 449, 521

\bibitem[Ibata et al.(2001)]{2001Natur.412...49I}
Ibata R. A., Irwin M. J., Lewis G. F., Ferguson A. M. N., Tanvir N. R., 
2001, Nat, 412, 49 

\bibitem[Kerins(2001)]{kerinsa}
Kerins E. J., 2001,
in Tr\^an~Thanh~Van J., Mellier Y., Moniez M., eds.,
Cosmological Physics with Gravitational Lensing.
EDP Sciences, Paris, p.43

\bibitem[Kerins et al.(2001)]{kerins}
Kerins E. J. et al., 2001, MNRAS, 323, 13

\bibitem[Kiraga \& Paczy\'nski(1994)]{kiraga}
Kiraga M., Paczy\'nski B., 1994, ApJ, 430, L101

\bibitem[Mann \& Whitney (1947)]{mann47}
Mann H., Whitney D., 1947, Ann. Math. Statist., 18, 50

\bibitem[Merritt \& Sellwood (1994)]{1994ApJ...425..551M}
Merritt D., Sellwood J. A., 1994, ApJ, 425, 551 

\bibitem[Paczy\'nski(1986)]{pac}
Paczy\'nski B., 1986, ApJ, 304, 1

\bibitem[Paulin-Henriksson et al.(2002)]{phtwo}
Paulin-Henriksson S. et al., 2002, ApJ, 576, L121

\bibitem[Paulin-Henriksson et al.(2003)]{phone}
Paulin-Henriksson S. et al., 2003, A\&A, in press (astro-ph/0207025)

\bibitem[Riffeser(2001)]{riff}
Riffeser A. et al., 2001, A\&A, 379, 362

\bibitem[Walterbos \& Kennicutt(1987)]{wk}
Walterbos R., Kennicutt R., 1987, A\&AS, 69, 311

\bibitem[Wielen et al.(1983)]{wielen} 
Wielen R., Jahrei\ss\ H., Kr\"uger R., 1983,
in Davis Philip A.G., Upgren A.R., eds., Proc.\ IAU Colloq.\ 76,
The Nearby Stars and the Stellar Luminsoity Function.
L. Davis Press, Schenectady, NY, p.163

\bibitem[White(1981)]{sdmw}
White S. D. M., 1981, MNRAS, 195, 1037

\end{thebibliography}
\end{document}